\documentclass{emulateapj}

\received{} \accepted{}

\shorttitle{Globular Cluster in M31}
\shortauthors{Ma et al.}

\begin{document}

\title{037-B327 in M31: Luminous Globular Cluster or Core of a Former Dwarf
Spheroidal Companion to M31?\altaffilmark{1}}

\author{
Jun Ma\altaffilmark{2}, Sidney van den Bergh\altaffilmark{3}, Hong
Wu\altaffilmark{2}, Yanbin Yang\altaffilmark{2}, Xu
Zhou\altaffilmark{2}, Jiansheng Chen\altaffilmark{2}, Zhenyu
Wu\altaffilmark{2}, Zhaoji Jiang\altaffilmark{2}, Jianghua
Wu\altaffilmark{2}}

\altaffiltext{1}{Based on observations made with the NASA/ESA
Hubble Space Telescope, obtained at the Space Telescope Science
Institute, which is operated by AURA, Inc., under NASA contract
NAS 5-26555. These observations are associated with proposal
9453.}

\altaffiltext{2}{National Astronomical Observatories, Chinese
Academy of Sciences, Beijing, 100012, P. R. China;
majun@vega.bac.pku.edu.cn}

\altaffiltext{3}{Dominion Astrophysical Observatory, Herzberg
Institute of Astrophysics, National Research Council, 5071 West
Saanich Road, Victoria, BC V9E 2E7, Canada}

\begin{abstract}

037-B327 is of interest because it is both the most luminous and
the most highly reddened cluster known in M31. Deep observations
with the Advanced Camera for Surveys on the $Hubble$ $Space$
$Telescope$ provide photometric data in the F606W band, and also
show that this cluster is crossed by a dust lane. We determined
the structural parameters of 037-B327 by fitting the observed
surface brightness distribution to a King model with
$r_c=0.72\arcsec(=2.69~\rm{pc})$, and
$r_t=5.87\arcsec(=21.93~\rm{pc})$, and a concentration index
$c=\log (r_t/r_c)=0.91$. The surface brightness profile appears to
be essentially flat within $0.25\arcsec$ of the center and shows
no signs of core collapse. Although the dust lane affects the
photometry, the King model fits the surface brightness profile
well except for the regions badly affected by the dust lane. We
also calculate the half-light radius
$r_h=1.11\arcsec(=4.15~\rm{pc})$. Combined with previous
photometry, we find that this object falls in the same region of
the $M_V$ versus log $R_h$ diagram as do $\omega$ Centauri, M54
and NGC 2419 in the Milky Way and the massive cluster G1 in M31.
All four of these objects have been claimed to be the stripped
cores of former dwarf galaxies. This suggests that 037-B327 may
also be the stripped core of a former dwarf companion to M31.

\end{abstract}

\keywords{galaxies: evolution -- galaxies: individual (M31) --
globular cluster: individual (037-B327)}

\section{Introduction}

It has been speculated that some of the most luminous known
globular clusters might be the remnants of tidally stripped dwarf
galaxies nuclei \citep{zinnecker88,freeman93,bassino94}. The study
of globular clusters in M31 was initiated by \citet{hubble32}, who
discovered 140 GCs with $m_{pg}\leq 18$ mag. The continued
importance of the study of GCs in this galaxy has been reviewed by
\citet{bh00}. M31 globular cluster B327 (B for `Baade') or Bo037
(Bo for `Bologna', see Battistini 1987), which, in the
nomenclature introduced by \citet{huchra91} will subsequently be
referred to as 037-B327. The extremely red color of this object
was first noted by \citet{kronmay60}.

The brightest globular clusters in M31 are more luminous than the
giant Galactic cluster $\omega$ Centauri. Among these are 037-B327
\citep{bergh68} and G1 \citep[see details from][]{bk02a}. The
latter has been considered as the possible remnant core of a
former dwarf galaxy which lost most of its envelope through tidal
interactions with M31 \citep{meylan97,meylan01}. Subsequently
\citet{mackey05} strengthened the \citet{meylan97} and
\citet{meylan01} conclusion.

In this paper, we have determined the structural parameters of
037-B327 using its deep image obtained with the Advanced Camera
for Survey (ACS) on the $Hubble$ $Space$ $Telescope$ $(HST)$.
Combined with the previous photometry, we find that this cluster
lies in the same region of the log $R_h$ versus $M_V$ diagram as
do $\omega$ Centauri, M54 and NGC 2419 in the Milky Way and G1 in
M31. This suggests that 037-B327 may also be the remnant core of a
now defunct dwarf companion to the Andromeda galaxy.

\begin{figure*}
\begin{center}
\centerline{\includegraphics[angle=0,width=120mm]{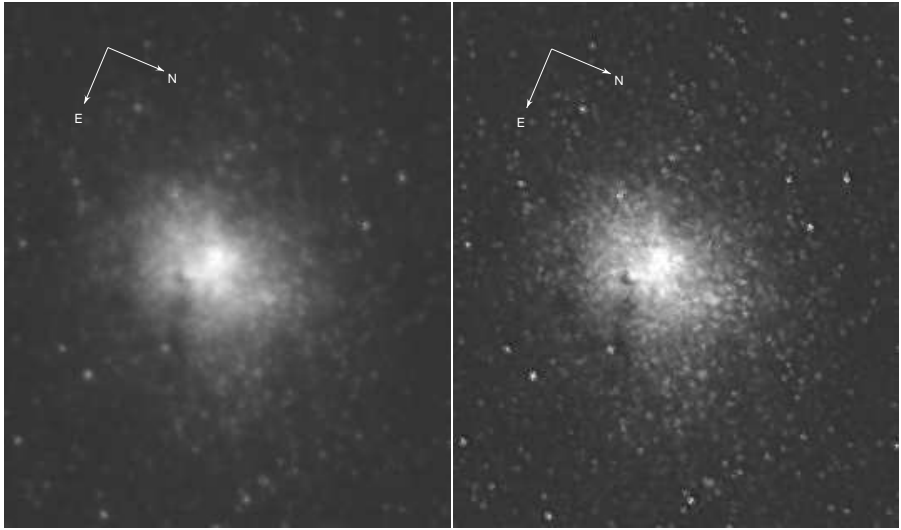}}
\caption{The image of GC 037-B327 observed in the F606 (left) and
its deconvolved counterpart (right). The central structure is
clearly more complex in the deconvolved image. The image size is
$7.8\arcsec\times8.8\arcsec$.} \label{fig1}
\end{center}
\end{figure*}

\section{Observations and Data Reduction}

We searched the $HST$ archive and found 037-B327 to have been
observed with the ACS-Wide Field Channel (WFC) in the F606W and
the F814W bands. We mainly used the image in F606W, which was
observed on 2004 August 2 with 2370.0 seconds of exposure time. We
deconvolve the image using the IRAF task Lucy \citep{lucy74}. The
image of 037-B327 observed in the F606 and its deconvolved
counterpart are shown in Figure 1. The central structure is
clearly more complex in the deconvolved image. The $HST$ ACS-WFC
resolution is $0.05\arcsec$ per pixel. We used the IRAF task
Ellipse to fit the image with a series of elliptical annuli from
the center to the outskirts, with the length of the semi-major
axis increasing by 10\% in each step. Figure 2 shows the
ellipticity and position angle, plotted as a function of the
semi-major axis. The ellipticity varies significantly with
position along the semi-major axis $a$. The mean ellipticity is
$\epsilon\simeq0.23$. The position angle P.A. is not significantly
variable for semi-major axis values $a$ larger than $0.5\arcsec$.
It is of interest to note that the high ellipticity of 037-B327,
which is the most luminous cluster in M31, confirms the empirical
rule \citep{bergh96} that the brightest globular cluster in a
galaxy is also usually one of the most flattened ones. The
ellipticities and position angles shown in Fig. 2 (particularly
near $r=0.5\arcsec$) are quite strongly affected by the dust lane.

\begin{figure}[htbp]
\figurenum{2} \epsscale{0.9}
\hspace{0cm}\rotatebox{-90}{\plotone{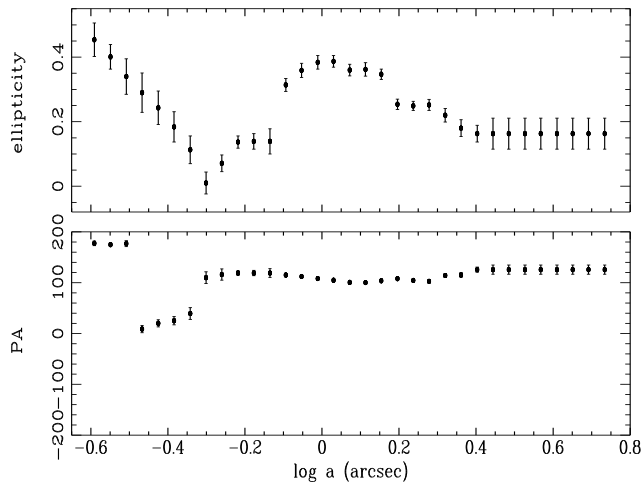}} \vspace{0.0cm}
\caption{Ellipticity and position angle as a function of the
semimajor axis, which are quite strongly affected by the dust
lane, particularly near $r=0.5\arcsec$ ($\log a=-0.3$).}
\label{fig2}
\end{figure}

We fitted King models \citep{king62} to the surface brightness
profiles. As usual, we parameterize the model with the core radius
$r_c$, the concentration index $c=\log (r_t/r_c)$ ($r_t$ is the
tidal radius.), and $\mu(0)$, the central surface brightness. The
derived parameters are: the core radius $r_c=0.72\arcsec$, the
tidal radius $r_t=5.87\arcsec$, implying the concentration index
$c=\log (r_t/r_c)=0.91$. The surface brightness profile appears to
be essentially flat within $0.25\arcsec$ of the center and shows
no signs of core collapse. The central surface brightness is
$17.21~\rm{mag} ~\rm{arcsec}^{-2}$. Figure 3 plots the surface
brightness profile and a fitted King model. As we noted, this
cluster contains a clear dust lane (see from the images in Figure
1, especially from the deconvolved image). So, some data points
are affected by this dust lane, which is evident in Figure 3.
Except for some photometric data affected badly by the dust lane,
the King model fits the surface brightness profile well. We also
calculate the half-light radius (the radius that contains half of
the light in projection) to be $r_h=1.11\arcsec$. With an adopted
a distance to M31 of 770 kpc \citep{meylan01}, the core radius,
the half-light radius and the tidal radius are $2.69~\rm{pc}$,
$4.15~\rm{pc}$ and $21.93~\rm{pc}$, respectively.

\begin{figure}[htbp]
\figurenum{3} \epsscale{1.1}
\hspace{0cm}\rotatebox{0}{\plotone{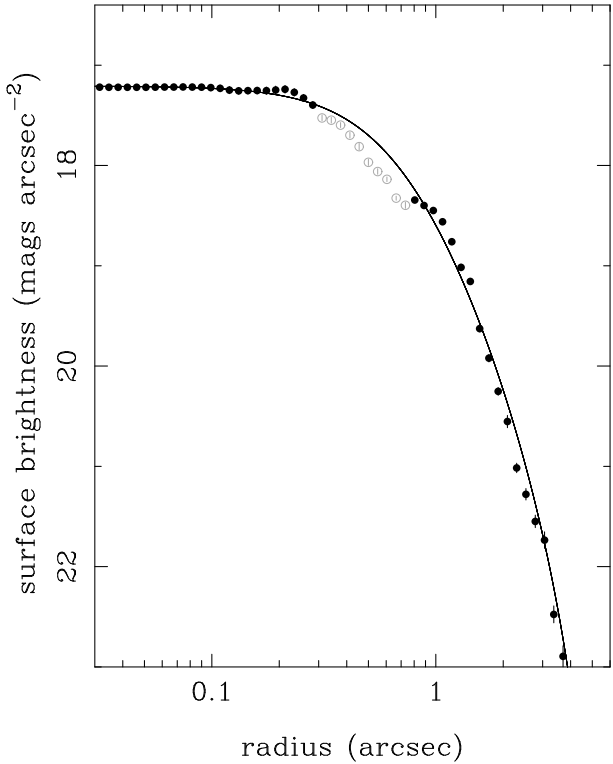}} \vspace{0.0cm}
\caption{Surface brightness profile of 037-B327, the continuous
line represents the King model fitted to the observed profile.
When fitting this profile we did not use the data plotted as open
circles since they were badly affected by the dust patches.}
\label{fig3}
\end{figure}

\section{Absolute Magnitude for 037-B327}

037-B327 is remarkable for being the most luminous and the most
highly reddened \citep{bergh68} cluster in M31. \citet{vete62a}
determined magnitudes of 257 M31 GC candidates in the $UBV$
photometric system including 037-B327. Using his photometric
catalog, \citet{vete62b} studied the intrinsic colors of M31 GCs,
and found that 037-B327 was the most highly reddened with
$E(B-V)=1.28$ in his sample of M31 GC candidates. Using
low-resolution spectroscopy, \citet{cram85} also found this
cluster to be the most highly reddened GC candidate in M31, with
$E(B-V)=1.48$. With a large database of multicolor photometry,
\citet{bh00} determined the reddening value for each individual
M31 GC including 037-B327 using the correlations between optical
and infrared colors and metallicity by defining various
``reddening-free'' parameters. Again, \citet{bk02a} derived the
reddening value for this cluster, using the spectroscopic
metallicity to predict the intrinsic colors. The dust lane showed
in Figure 3, might be responsible for the bulk of the reddening.
In this paper, we adopted $E(B-V)=1.32\pm0.05$ for 037-B327
derived by \citet{bk02a} by the weighted combination of values
from their two methods \citep[see details from][]{bk02a}, and
$m_V=16.82$ presented by \citet{bh00}. Assuming $R_V=3.1$ and a
distance to M31 of 770 kpc \citep{meylan01}, the absolute
magnitude of 037-B327 is $M_V=-11.71$, which makes it the most
luminous globular cluster in M31.

\begin{figure}[htbp]
\figurenum{4} \epsscale{0.9}
\hspace{0cm}\rotatebox{-90}{\plotone{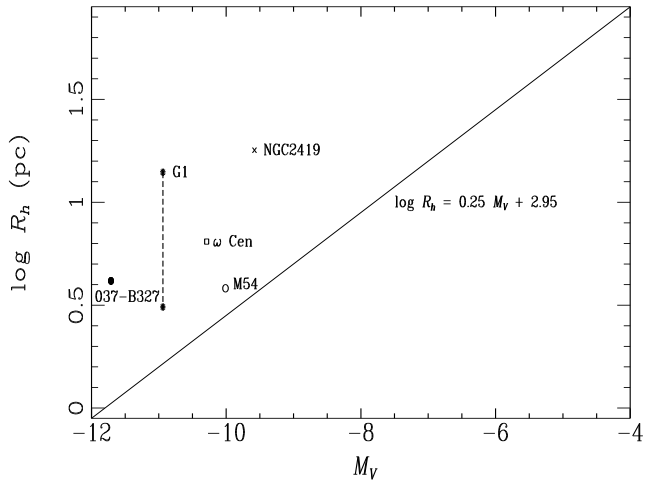}} \vspace{0.0cm}
\caption{Relation between $M_{V}$ and $R_{h}$ for 037-B327. The
figure shows that 037-B327 lies above and to the brightward of
Equation (1) in the $M_{V}$ versus $\log R_{h}$ plot. The data for
$\omega$ Centauri, M54, NGC 2419, and G1 were taken from
\citet{mackey05}. G1 is marked by two linked stars - representing
the two discrepant measurements of $R_h$ for this cluster. The
upper point is the measurement of \citet{meylan01}, while the
lower point is that of \citet{bk02b}}\label{fig4}
\end{figure}

\section{Luminous Globular Cluster or Core of a Former Dwarf
Spheroidal Companion to M31}

As a globular cluster evolves its core contracts and its envelope
expands. However, it has been shown by \citet{spitzer92},
\citet{henon73}, \citet{lightman78}, and \citet{murphy90} that the
half-light radius of an evolving cluster changes little over
periods as long as 10 relaxation times. The half-light radius of a
cluster therefore can be used to trace the initial size of a
cluster, and hence the physical conditions in its host galaxy at
early epochs. In previous papers \citep{bergh04,mackey05}, it was
showed that three Galactic globular clusters and one M31 globular
cluster lie above and to the brightward of the line

\begin{equation}
\log R_{h} = 0.25 M_{V} + 2.95,
\end{equation}
where $R_{h}$ and $M_{V}$ are the half-light radius and the
absolute magnitude of a globular cluster. Figure 11 of
\citet{mackey05} showed that the overwhelming majority of normal
globular clusters lie below and to the faintward of the line
defined by Equation (1). The clusters that do fall above the
relation defined by Equation (1) are mostly objects suspected of
being the cores of now defunct dwarf galaxies.

Figure 4 shows a plot of $\log R_{h}$ versus $M_{V}$. On this plot
037-B327 is seen to lie above and to the brightward of the line
defined by Equation (1) as do $\omega$ Centauri, M54 and NGC 2419
in the Milky Way and the massive cluster G1 in M31. All four of
these objects have been suggested to be the stripped cores of
former dwarf galaxies \citep[for details see][]{mackey05}. This
result suggests that the most luminous cluster 037-B327 in M31 may
also be the stripped core of a now defunct dwarf companion to M31.

\section{Summary}

In this paper, we determine the structural parameters of 037-B327
that were derived from an F606W image that was obtained with the
Advanced Camera for Surveys on the $Hubble$ $Space$ $Telescope$,
by fitting between the surface brightness distribution and the
King model. Combined with the previous photometry, we find that
this object falls in the same region of the $M_V$ versus $R_h$
diagram as do $\omega$ Centauri, M54 and NGC 2419 in the Milky Way
and the massive cluster G1 in M31 on the size (log $R_h$) versus
luminosity ($M_V)$ diagram. All four of these objects have been
suggested to be the stripped cores of former dwarf galaxies. So,
we argue that 037-B327 may also be the core of a former dwarf
spheroidal companion to M31. We also compared the images of the
F606W and F814W, and did not find any difference in the colors of
the brightest incipiently resolved stars, where this term is used
in the sense that the image is not clearly resolved into
individual stars, but has a mottled or granular appearance, which
was employed by \citet{baade63}.

\acknowledgments We would like to thank the first anonymous
referee for catching our error in the M31 image scale. We are also
indebt to the second anonymous referee for his/her insightful
comments and suggestions that improved this paper greatly. This
work has been supported by the Chinese National Key Basic Research
Science Foundation (NKBRSF TG199075402) and by the Chinese
National Natural Science Foundation, No. 10473012, 10573020, and
10333060.

\end{document}